\documentclass[final]{svjour2}
\usepackage{graphicx}
\usepackage{rotating}
\usepackage{amssymb}
\usepackage{mathptmx}
\usepackage[numbers]{natbib}
\usepackage{amsmath}
\usepackage{upgreek}
\makeatletter
\journalname{Journal of Low Temperature Physics}

\bibpunct{}{}{,}{s}{}{,}

\begin{document}

\newcommand{\hdblarrow}{H\makebox[0.9ex][l]{$\downdownarrows$}-}
\title{Analysis of impedance and noise data of an X-ray transition-edge sensor using complex thermal models}

\author{M. R. J. Palosaari$^1$, K. M. Kinnunen$^1$, M. L. Ridder$^2$, J. van der Kuur$^2$, H. F. C. Hoevers$^2$ and I. J. Maasilta$^1$}

\institute{1:Nanoscience Center, Department of Physics, P. O. Box 35, FI-40014 University of Jyv\"askyl\"a, Finland\\
\email{mikko.palosaari@jyu.fi}
\\2:SRON Netherlands Institute for Space Research, Sorbonnelaan 2, 3584 CA Utrecht, the Netherlands }

\date{21.11.2011}

\maketitle


\begin{abstract}
	The so-called excess noise limits the energy resolution of transition-edge sensor (TES) detectors, and its physical origin has been unclear, with many competing models proposed. Here we present the noise and impedance data analysis of a rectangular X-ray Ti/Au TES fabricated at SRON. To account for all the major features in the impedance and noise data simultaneously, we have used a thermal model consisting of three blocks of heat capacities, whereas a two-block model is clearly insufficient. The implication is that, for these detectors, the excess noise is simply thermal fluctuation noise of the internal parts of the device. Equations for the impedance and noise for a three-block model are also given.
	
	\end{abstract}
\section{Introduction}
Calorimeters and bolometers based on superconducting transition-edge sensors (TES) have proven to be valuable tools in a number of applications in a broad energy range \cite{irwin}. The number of pixels in TES detector arrays are constantly increasing and the limits of single pixel performance are being pushed closer to the theoretical limits. However, the lack of understanding of some of the noise components ("excess noise") has plagued the field in recent years  \cite{ullom1, kimmo1}. One source for the excess noise, in addition to the recently introduced non-equilibrium Johnson noise \cite{kent}, could be a complex thermal circuit of the device: if the device consists of several blocks of heat capacity, more thermal fluctuation noise will exist \cite{hoevers, kimmouus}. To determine unambiguously the thermal noise components of the device, the thermal circuit should be determined independently. This can be done by measuring the complex impedance in addition to the noise \cite{lindemann}. For the device discussed here, the impedance data is fit well by a three-block thermal model, and the model simultaneously explains all the noise of the device. Thus, all noise sources are fully understood in the detector types presented here. 
\section{Theoretical models}
\begin{figure}[ht]
\begin{center}
\includegraphics[width=0.9\linewidth,keepaspectratio]{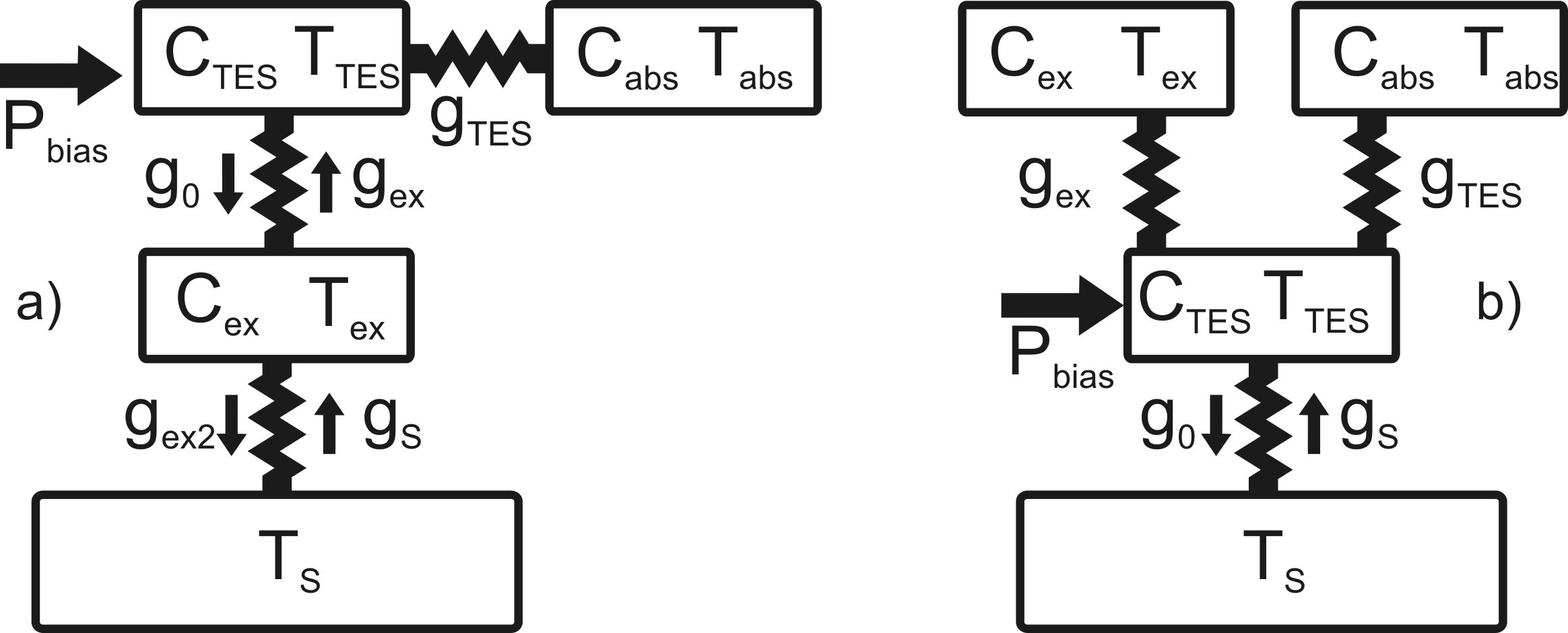}
\end{center}
\caption{ a) A model with one hanging and an intermediate block between the TES and the bath (IH model). b)   A model with two hanging thermal blocks (2H model).}
\label{block}
\end{figure}

To fully characterize the thermal and electrical properties of many TES devices, the conventional thermal model of one heat capacity connected to a heat bath is too simple. Even the  model with one additional thermal block does not always fully fit the measured noise and impedance \cite{AIP,IEEE}. Here we have used a system of three thermal blocks. We have analyzed the measured data with two different variations: one with both a hanging and an extra intermediate block between the TES and the heat bath, and another with two hanging thermal blocks (Fig. \ref{block}).  
The derivation and full theoretical discussion of the impedance and noise of the three-block models will be presented elsewhere \cite{ilari}, here we only cite the results for the IH model (Fig. \ref{block} a). Equations for 2H model (Fig. \ref{block} b) are similar.  The IH model assumes that heat flows first from the TES to the ''intermediate'' or excess heat capacity $C_{ex}$, and only then to the heat bath, therefore the steady state temperatures of the TES and the excess heat capacity are not equal. The hanging heat capacity $C_{abs}$ could represent the absorber, but does not in general need to do so, and in steady state it has the same temperature as the TES. There are now five different dynamical thermal conductances that need to be defined, one between the TES and the ''absorber'' , two between the TES and the excess heat capacity and two between the excess heat capacity and the bath:
$g_{TES}=nAT_{TES}^{n-1}$, $ g_0=mBT_{TES}^{m-1}$, $g_{ex}=mBT_{ex}^{m-1}$, $g_{ex2}=pCT_{ex}^{p-1}$, $g_s=pCT_s^{p-1}$,
where the power between the TES and the absorber, the TES and the excess heat capacity, and the excess heat capacity and the bath are given by  $P=A(T_{TES}^n-T_{abs}^n)$,  $P=B(T_{TES}^m-T_{ex}^m)$    and   $P=C(T_{ex}^p-T_{s}^p)$,   respectively (see Fig. 1).
\subsection{Complex Impedance}
The complex impedance of the model (Fig. 1 a) is
\begin{equation}
Z_{TES}=R_0(1+\beta) +\frac{\cal L}{1-{\cal L}}\frac{R_0(2+\beta)}{1+i\omega\tau_I-A(\omega)-B(\omega)},
\end{equation}
where   
$A(\omega) = \frac{1}{1-{\cal L}}\frac{g_{tes}}{(g_{tes}+g_0)}\frac{1}{1+i\omega\tau_{abs}}$, 
$B(\omega) = \frac{1}{1-{\cal L}}\frac{g_{0}g_{ex}}{(g_{tes}+g_0)(g_{ex}+g_{ex2})}\frac{1}{1+i\omega\tau_{ex}}$,\\
${\cal L} =P_0\alpha/[(g_{tes}+g_0)T_0]$, $\tau_I =C_{tes}/[(g_{tes}+g_0)(1-{\cal L})]$, $ \tau_{abs} =C_{abs}/g_{tes}$ and \\ $ \tau_{ex} =C_{ex}/(g_{ex}+g_{ex2})$,
and the transition steepness parameters are  \\  $\alpha=\partial\log(R)/\partial\log(T)$,   $\beta=\partial\log(R)/\partial\log(I)$   at the bias point   $R_0$   (with power   $P_0$   and temperature   $T_0$).
\subsection{Noise}
Three major classes of unavoidable noise sources are included: the power fluctuations in the thermal circuit, the electrical thermal noise of the detector (Johnson noise), and the Johnson noise of the shunt resistor. We disregard correlations between fluctuations in the thermal conductances. 

The frequency dependent current responsivity,   $s_I(\omega)=I_\omega/P_\omega$,   for power input in  the TES heat capacity   $C_{tes}$   can be written \cite{ilari} as  
\begin{equation}
s_I(\omega)=-\frac{1}{Z_{circ}I_0}\frac{Z_{TES}-R_0(1+\beta)}{R_0(2+\beta)},
\end{equation}
where $Z_{circ}= Z_{TES}+R_L+i\omega L$, $R_L$ is the Thevenin equivalent circuit resistance (shunt+parasitic) and $L$ is the circuit inductance.

Now the thermal fluctuation current noise terms (one phonon noise term and two internal thermal fluctuation noise (ITFN) terms) are
 \begin{align}
|I|^2_{ph}&=P_{ph}^2|s_I(\omega)|^2\frac{g_{ex}^2}{(g_{ex}+g_{ex2})^2}\frac{1}{1+\omega^2\tau_{ex}^2},\\
|I|^2_{ITFN,1}&=P_{tes}^2|s_I(\omega)|^2\frac{\omega^2\tau_{abs}^2}{1+\omega^2\tau_{abs}^2},\\
|I|^2_{ITFN,2}&=P_{ex}^2|s_I(\omega)|^2\frac{g_{ex2}^2/(g_{ex2}+g_{ex})^2+\omega^2\tau_{ex}^2}{1+\omega^2\tau_{ex}^2},\\
\end{align}
where $P_{ph}^2= 2k_B (T_{ex}^2g_{ex2}+T_s^2g_s)$, $P_{tes}^2= 4k_BT_{0}^2g_{tes}$ and $P_{ex}^2= 2k_B (T_{0}^2g_{0}+T_{ex}^2g_{ex})$.

The non-equilibrium Johnson current noise in the TES film is given by
\begin{equation}
|I|^2_{J}=\frac{V_\omega^2}{ |Z_{circ,\infty}+\frac{{\cal L}(R_0-R_L-i\omega L)}{1+i\omega\tau_{tes}-(1-{\cal L})(A(\omega)+B(\omega))}|^2},
\end{equation}
where \cite{kent} $V_\omega^2=4k_BT_0R_0(1+2\beta)$  and $Z_{circ,\infty}=R_0(1+\beta)+R_L+i\omega L$.
The Johnson noise due to the shunt and parasitic resistances is simply
$|I|^2_{sh}=V_{\omega,sh}^2/|Z_{circ}|^2$,
with 
$V_{\omega,sh}^2=4k_BT_{sh}R_L$ if both the parasitic resistance and the actual shunt are at temperature $T_{sh}$.

\section{Experiments and Analysis}
\begin{figure}[ht!]
\begin{center}
\includegraphics[width=1\linewidth,keepaspectratio]{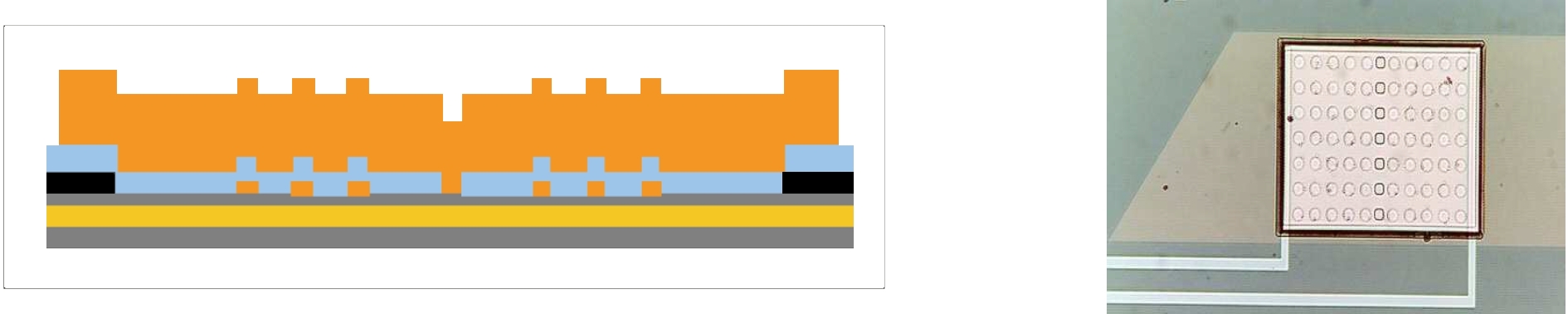}
\end{center}
\caption{ (Color online) Left: Schematic cross-sectional view of the device. Orange: Cu absorber and dots, blue: SiOx insulator, gray: Ti layers, yellow: Au layer, black: Nb contacts. Right: Optical micrograph of the TES. The size of the Ti/Au TES is 186 x 150 $\upmu$m$^2$ and it is on a 1 $\upmu$m thick SiN membrane fabricated on Si (110) surface.}
\label{tes}
\end{figure}

The measured TES was a pixel from an X-ray array fabricated at SRON (Fig. \ref{tes}). 
It features a 206 x 162 x 1 $\upmu$m$^3$ Cu absorber on top of a SiOx insulator that is coupled to the TES
through seven rectangular vias at the center of the TES. There are also 10 x 7  Cu dots of 10 $\upmu$m diameter on top of the TES
film (but not in contact with the absorber) to tune the transition properties.   The critical temperature, $T_C$, is 125.5 mK and normal state resistance $R_N \sim$ 300 m$\Omega$. Measurements were performed in a compact plastic dilution refrigerator at Jyv\"askyl\"a, with an old NIST SQUID readout \cite{cher} and a FLL electronics unit designed at SRON. The complex impedance was measured up to 100 kHz, taking into account of the transfer function of the readout circuit \cite{sron}. 

In the analysis, the measured complex impedance and noise were fitted simultaneously by eye with the equations described above, as high-dimensional non-linear
least-squares fitting would be demanding to implement. The heat capacities, $g_0$, $g_{TES}$ and $T_{ex}$ were free parameters. The other thermal conductances were calculated using the constraint that the total thermal conductance to the bath is fixed by the I-V data and that the links on both sides of the intermediate block have
the same thermal exponent $p=m$. This is physically reasonable if all the conductances are dominated by the SiN membrane. During the fitting four curves were plotted on top of the experimental data: The noise as a function of frequency, the complex impedance and also the real and imaginary parts of the impedance as a function of frequency. White noise of 4 pA/$\sqrt{\mathrm{Hz}}$ for the SQUID was included in the noise analysis. 
\section{Results and Discussion}
\begin{figure}[ht!]
\begin{center}
\includegraphics[width=1\linewidth,keepaspectratio]{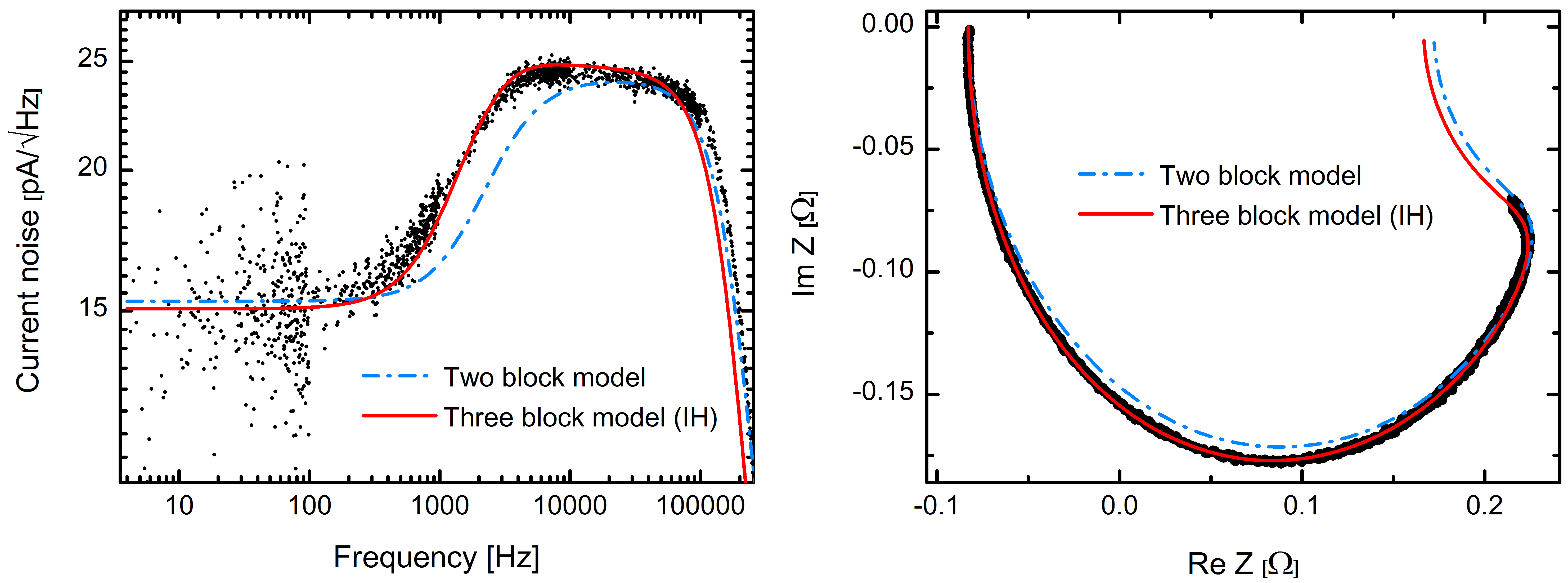}
\end{center}
\caption{ (Color online) Comparison between a two-block model and the IH three-block model at 20\% R/R$_N$.}
\label{comp}
\end{figure}\begin{figure}[hm!]
\begin{center}
\includegraphics[width=1\linewidth,keepaspectratio]{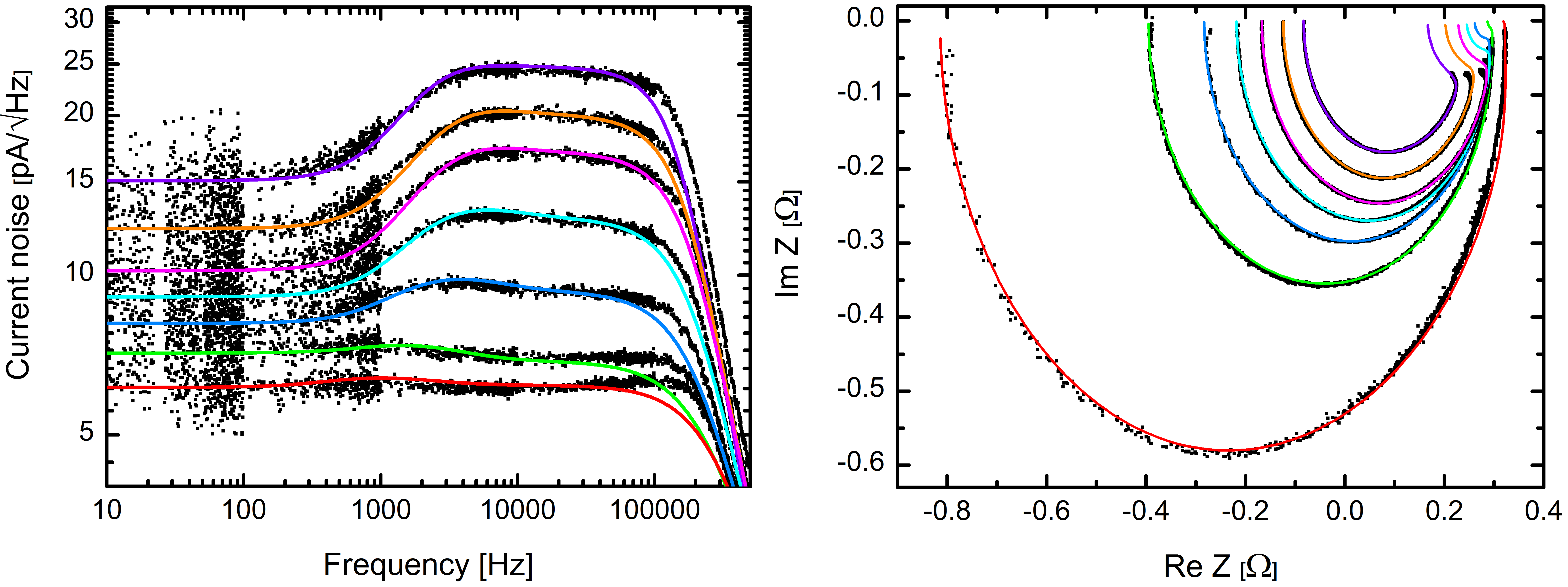}
\end{center}
\caption{ (Color online) Left: the measured and fitted current noise of the device. Right: the measured and fitted complex impedance of the TES. Bias points between 20\%-80\% $R_N$ in 10\% intervals are presented. The higher bias points correspond to the lower noise levels and bigger semicircles. Only IH model fits are shown, 2H fits look identical.}
\label{data}
\end{figure}

\begin{figure}[ht!]
\begin{center}
\includegraphics[width=1\linewidth,keepaspectratio]{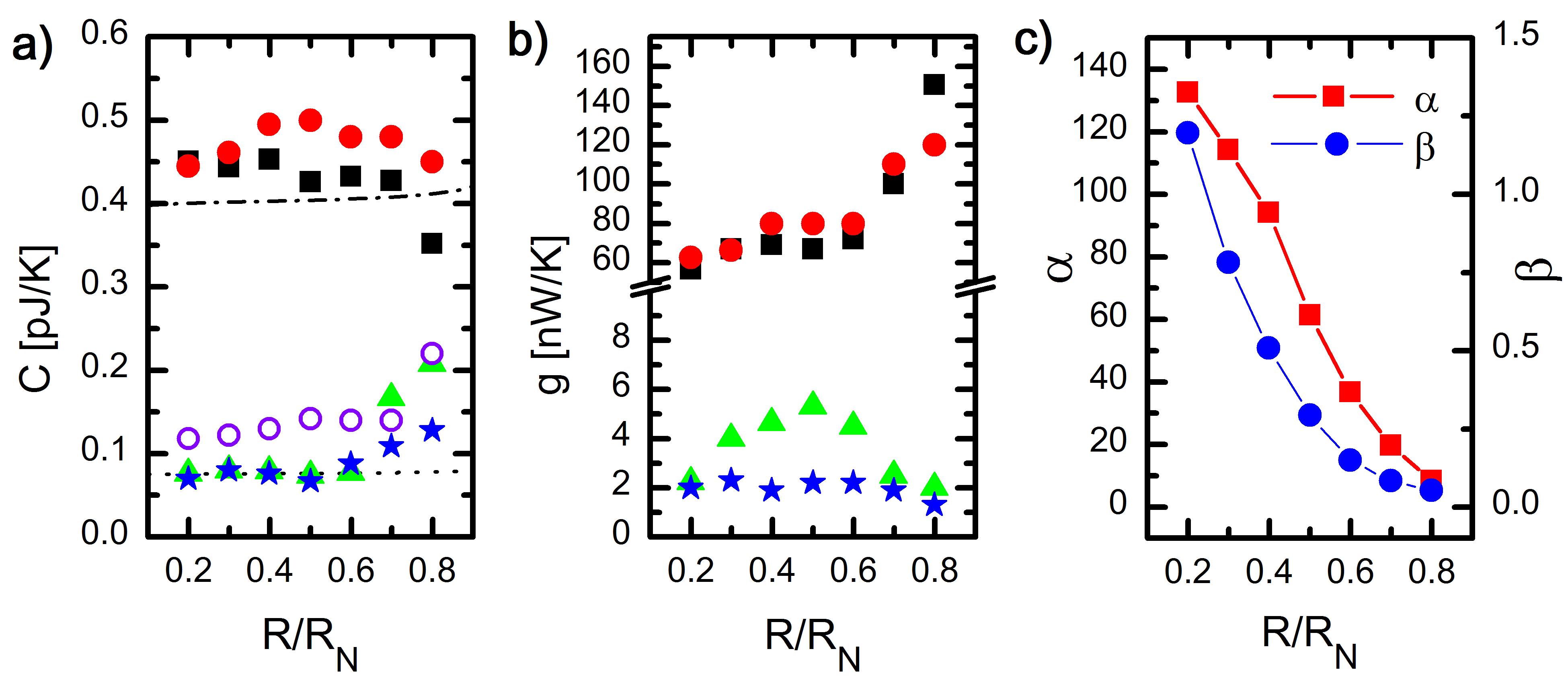}
\end{center}
\caption{ (Color online) Some of the parameters from both IH and 2H models vs. bias point. a) $C_{TES}$ (IH triangles, 2H stars), $C_{abs}$ (IH squares, 2H circles) and $C_{ex}$ (IH open circles) and the estimated theoretical values as dashed lines.  b) $g_{TES}$ (IH squares, 2H circles), $g_{0}$ (IH triangles) and $g_{Ex}$ (2H stars). c) $\alpha$ and $\beta$. }
\label{fitpar}
\end{figure}

One can fit all the major noise and impedance features with both three-block models (IH and 2H, Fig. 1), whereas a simpler two-block model (TES + hanging $C$) cannot produce an adequate fit, as can be seen in Fig. \ref{comp}. In Fig. \ref{data} the measured data and fitted theoretical curves for the IH model are presented. In the noise data, one can see that there is some deviation with the measured and the fitted data above 100 kHz. This is likely because the transfer function of the circuit was not corrected for in the noise data. 
We also analysed the measured data with the two hanging (2H) block model of Fig. \ref{block} b) and obtained qualitatively identical goodness of fit results with the IH model. The fitted parameter values for both models are plotted in Fig. \ref{fitpar}, except for $C_{ex}$ in the 2H model which was approximately constant $C_{ex} \approx 0.08$ pJ/K. 
We also estimated the theoretical heat capacity of the Cu absorber, $C_{abs}=\gamma T_{TES}$, to be around 0.40 pJ/K, which is quite close to the fitted $C_{abs}$, as shown in Fig. \ref{fitpar} a). The TES film heat capacity $C_{TES}$ was also calculated with a  correction \cite{kozo} to the BCS heat capacity jump for bilayer films, to get values around 0.08 pJ/K, which again agrees with our results.  We see that the two model variants do not produce significant differences, and cannot be easily differentiated from each other. We do not want to draw conclusions from the bias dependence of the parameters yet, because we have evidence that the IH model is required even in devices without an absorber  \cite{kimmouus}.

In conclusion, good simultaneous fit to complex impedance and noise can only be achieved with a three-block model in these devices. The fitting parameter values are consistent with the heat capacities being the TES film, the absorber and a third unknown heat capacity of the order of 0.1 pJ/K. No unexplained excess noise remains after the three-block fit.

\begin{acknowledgements}
This work was supported by ESA contract no AO/1-4005/01/NL/HB, the Finnish Funding Agency for Technology and Innovation TEKES and EU through the regional funds, and the Finnish Academy project no. 128532. M. P. would like to thank the National Graduate School in Materials Physics for funding.
\end{acknowledgements}


\begin{thebibliography}{99}

\bibitem{irwin}
K. Irwin and G. Hilton, in  {\it Cryogenic Particle Detection} edited by. Ch. Enss, Springer, Berlin, 63 (2005).

\bibitem{ullom1}
J. N. Ullom, {\it et al.}, 
Appl. Phys. Lett. 
\textbf{84}, 4206 (2004).

\bibitem{kimmo1}
K. M. Kinnunen, A. K. Nuottaj\"arvi, J. Lepp\"aniemi, and I. J.
Maasilta, J. Low Temp. Phys. \textbf{151}, 119 (2008).

\bibitem{kent}
K. D. Irwin, Nucl. Instrum. Meth. A \textbf{559}, 718 (2006).

\bibitem{hoevers} 
H. F. C. Hoevers, A. C. Bento, M. P. Bruijn, L. Gottardi, M. A. N. Korevaar, W. A. Mels, and P. A. J. de Korte, Appl. Phys. Lett. \textbf{77}, 4422 (2000).

\bibitem{kimmouus} 
K. M. Kinnunen, M. R. J. Palosaari, and I. J. Maasilta submitted for publication, arxiv:1111.4098v1.

\bibitem{lindemann}
M. Lindeman, {\it et al.},
Rev. Sci. Instrum. \textbf{75}, 1283 (2004).

\bibitem{AIP}
I. J. Maasilta and K. M. Kinnunen, AIP Conf. Proc. \textbf{1185}, 38 (2009).

\bibitem{IEEE}
Y. Zhao, {\it et al.},
IEEE Trans. Appl. Supercond. \textbf{21}, 227 (2011).

\bibitem{ilari}
I.J. Maasilta, to be published.

\bibitem{cher}
J. A. Chervenak, {\it et al.},
 Appl. Phys. Lett. {\bf74}, 4043 (1999).

\bibitem{sron}
Y. Takei, {\it et al.},
J. Low Temp. Phys. \textbf{151}, 161 (2008).


\bibitem{kozo}
A. Kozorezov {\it et al.},
\textbf{1185}, 27 (2009).

\end{thebibliography}
\end{document}